\documentclass[longbibliography,twocolumn,aps,prb,showkeys,superscriptaddress,letterpaper]{revtex4-2}
\usepackage[utf8]{inputenc}
\usepackage{amsmath}
\usepackage{amssymb}
\usepackage{bbold}
\usepackage{graphicx}
\usepackage{lipsum}
\graphicspath{{plots/}}
\usepackage{color}

\usepackage[unicode=true,
 bookmarks=true,bookmarksnumbered=false,bookmarksopen=false,
 breaklinks=false,pdfborder={0 0 0},pdfborderstyle={},backref=false,colorlinks=true,]
 {hyperref}
\hypersetup{linkcolor=blue,citecolor=blue,urlcolor=blue}

\begin{document}

\title{Dissipation effects in the Su-Schrieffer-Heeger model coupled to a metallic environment}

\author{Leandro M. Arancibia}

\affiliation{%
Instituto Interdisciplinario de Ciencias Básicas (ICB-CONICET), Universidad Nacional de Cuyo, Padre Jorge Contreras 1300, Mendoza 5502, Argentina
}
\affiliation{Facultad de Ciencias Exactas y Naturales, Universidad Nacional de Cuyo, Padre Jorge Contreras 1300, Mendoza 5502, Argentina
}

\author{Cristián G. Sánchez}

\affiliation{%
Instituto Interdisciplinario de Ciencias Básicas (ICB-CONICET), Universidad Nacional de Cuyo, Padre Jorge Contreras 1300, Mendoza 5502, Argentina
}
\affiliation{Facultad de Ciencias Exactas y Naturales, Universidad Nacional de Cuyo, Padre Jorge Contreras 1300, Mendoza 5502, Argentina
}

\author{Alejandro M. Lobos}\email{alejandro.martin.lobos@gmail.com}

\affiliation{%
Instituto Interdisciplinario de Ciencias Básicas (ICB-CONICET), Universidad Nacional de Cuyo, Padre Jorge Contreras 1300, Mendoza 5502, Argentina
}
\affiliation{Facultad de Ciencias Exactas y Naturales, Universidad Nacional de Cuyo, Padre Jorge Contreras 1300, Mendoza 5502, Argentina
}

\begin{abstract}
We theoretically study the electronic and lattice properties of a 
trans-polyacetylene (tPA) molecule deposited on top of a metallic substrate at equilibrium. We describe the system using a modified Su-Schrieffer-Heeger (SSH) model generalized to incorporate the effects of a metallic environment, represented by independent one-dimensional semi-infinite chains coupled to each site of the SSH chain (i.e., ``local bath approximation"). We focus on the zero-temperature case and obtain the physical properties of an $N$-site tPA chain deposited on a metallic surface by minimizing its total ground-state energy (i.e., electronic plus lattice degrees of freedom) as a function of the $N$ lattice-site positions. Interestingly, in the case of a homogeneous metallic substrate, where all coupling parameters are assumed identical, the SSH chain undergoes a zero-temperature  insulator-to-metal transition as the coupling parameter $\gamma_0$ reaches a critical value where the Peierls dimerization is fully suppressed and the system becomes metallic. In addition, our model can be generalized to describe  inhomogeneous situations where the substrate contains metallic and insulating regions, as usually occurs in realistic experiments containing accidentally oxidized decoupling layers. In this case, our results predict the occurrence  of local nucleation of the metalized or the Peierls-dimerized phase within the same tPA molecule, depending on whether the surface directly beneath the molecule is metallic or insulating, respectively. We finally discuss the relevance of our findings for both the correct interpretation of existing tPA/Cu(110) experiments, as well as for their possible utility in the design of novel organic nanoelectronic devices. 
\end{abstract}

\maketitle

\section{Introduction}\label{sec:intro}
Conducting polymers emerged as a transformative class of materials following the discovery of trans–polyacetylene (tPA), widely regarded as the simplest precursor of this family~\cite{heeger2001nobel}. This polymer exhibits remarkable electronic properties, including the ability to enhance its conductivity by several orders of magnitude through chemical or electrostatic doping~\cite{moraes1985first}, as well as the capacity to host non-linear excitations such as solitons that may carry charge without spin or, conversely, spin without charge. Shortly after the first experimental syntheses of tPA, Su, Schrieffer, and Heeger (SSH) introduced a tight-binding Hamiltonian that successfully captured its essential physics, providing a unified explanation for the Peierls gap, bond dimerization, and the emergence of topological soliton excitations \cite{Su79_Solitons_in_polyacetylene, Heeger88_Solitons_in_conducting_polymers}. 

Nearly five decades later, both tPA and the SSH model continue to motivate extensive research across condensed matter physics and molecular electronics. Recent advances include the first controlled synthesis of individual tPA chains on hybrid metal–insulator surfaces~\cite{Wang19_Solitons_in_individual_PA_molecules}, establishing a critical experimental platform for the development of organic nanoscale devices. On the theoretical front, several works have explored the potential of tPA for soliton-mediated charge transport~\cite{HernangomezPerez20_Solitonics_with_PA}, the engineering of domain walls supporting quantized charge accumulation~\cite{Arancibia25_Towards_electrical_DW_control}, and the encoding or transfer of information in molecular architectures~\cite{kim2017switching}. Together, these studies underscore the enduring relevance of tPA both as  a model system as well as a versatile building block for next-generation quantum and organic nanotechnologies \cite{Farchioni_Grosso_Organic_electronic_metals_book, Farchioni_Organic_Electronic_Materials}.

A natural question that arises in this context is the following: what happens to the characteristic properties of tPA when the polymer is coupled to a metallic substrate? Although the SSH model has been extensively validated for describing the topological features and solitonic excitations of tPA under idealized, isolated conditions, its behavior in contact with more realistic environments, where the chain interacts with a surface, a reservoir, or structural disorder, remains comparatively unexplored. Such coupling can substantially modify the electronic structure, destabilize Peierls-like configurations, or even alter the very nature of localized excitations, raising fundamental questions about the persistence or suppression of bond dimerization once the system is no longer strictly isolated. Extending traditional models to incorporate these additional degrees of freedom is therefore essential for assessing the limits of validity of the SSH framework and for developing more realistic descriptions of molecular systems in contact with conducting or insulating substrates \cite{edalatmanesh2025non}.

\begin{figure}
\centering
\begin{minipage}[t]{0.35\textwidth}
    \raggedright
    \text{(a)}\\[1.5cm]
    \includegraphics[width=\textwidth]{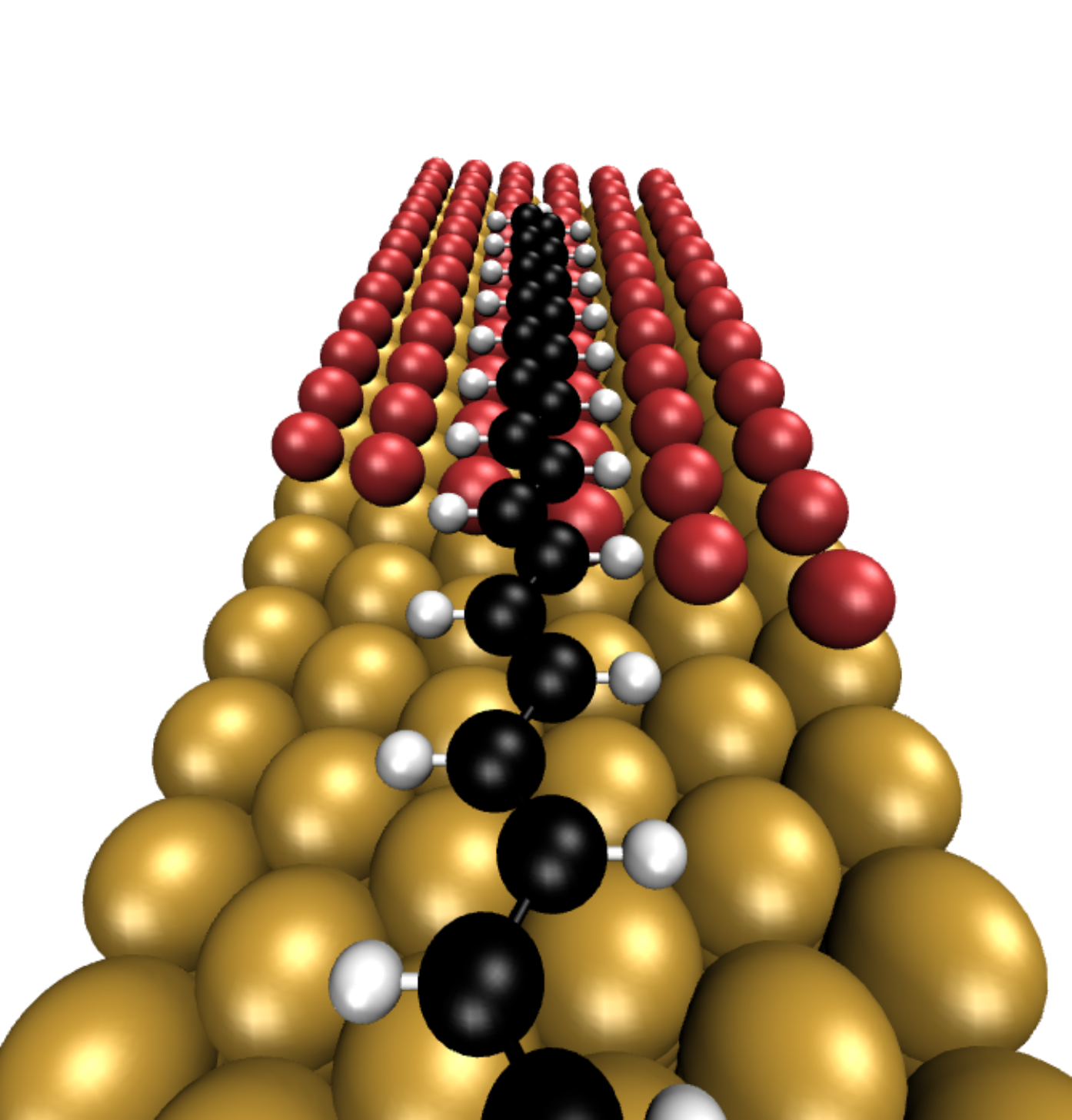}
\end{minipage}
\begin{minipage}[t]{0.6\textwidth}
\vspace{0.5cm}
    \raggedright
    \text{(b)}\\
    \includegraphics[width=0.7\textwidth]{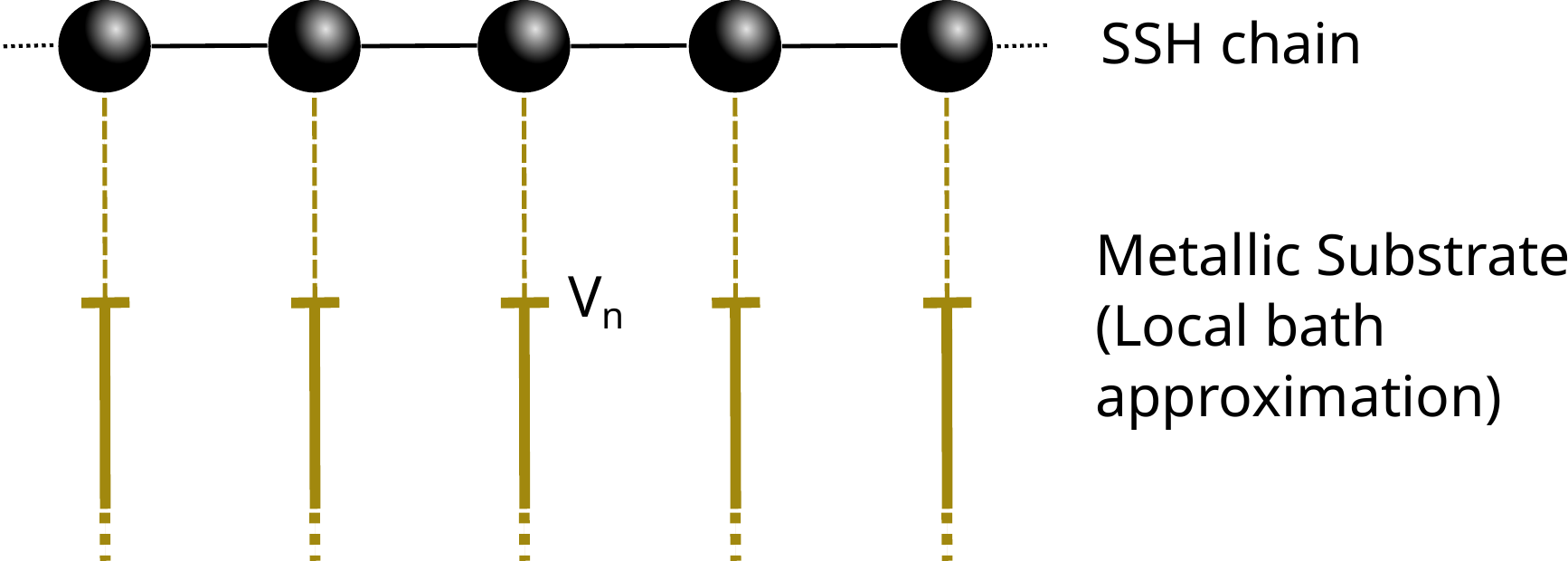}
\end{minipage}
    \caption{(Color online) (a) Schematic representation of a tPA molecule deposited on a hybrid surface composed of a metallic region (e.g., Cu(110)) and an insulating region (e.g., CuO). (b) Representation of the dissipative SSH model, where each site of the chain is coupled to a local reservoir (i.e., a semi-infinite vertical chain) via the coupling parameter $V_n$.
}
    \label{fig:onsurface_tPA_scheme}
\end{figure}

Within this broader context, we now formulate the central research problem addressed in this work. Specifically, we investigate a $N-$site tPA chain coupled to a metallic or hybrid substrate (see Fig.~\ref{fig:onsurface_tPA_scheme}) where the lattice degrees of freedom are self-consistently determined from the minimization of the total free energy, explicitly incorporating substrate-induced effects. Our theoretical analysis contains two original elements. First, we construct an extended Hamiltonian based on the SSH framework that enables a more realistic description than is typically considered, by explicitly including both the ionic coordinates and the influence of the surface. This formulation allows us to generate quantitative predictions for prospective experiments on a broad class of systems beyond tPA that can be mapped onto our Hamiltonian. Second, our approach contributes to the ongoing discussion surrounding recent experimental evidence, particularly the results reported by Wang \emph{et al.}, by providing a complementary theoretical perspective grounded in a fully relaxed and substrate-sensitive model \cite{Wang19_Solitons_in_individual_PA_molecules}.

\section{Theoretical model}
We describe the tPA molecule deposited ontop of a metallic surface with the following model:
\begin{align}
H&=H_\text{SSH}+H_\text{sub}+H_\text{mix},\label{eq:H}\\
H_\text{SSH}&=\sum_{n=1,\sigma}^{N} \left[-t_0-\alpha \left(u_{n+1}-u_n\right)\right]\left(c_{n+1,\sigma}^\dagger c_{n,\sigma}+ \text{H.c.}\right)\nonumber \\
&+\frac{K}{2}\sum_{n=1}^{N} \left(u_{n+1}-u_n\right)^2,\label{eq:H_SSH}\\
H_\text{sub}&=\sum_{n=1,\sigma}^{N}\sum_{j=1}^\infty \left[\mu_r d_{n,\sigma,j}^\dagger d_{n,\sigma,j} - t_{r} \left(d_{n,\sigma,j+1}^\dagger d_{n,\sigma,j} +\text{H.c.}\right)\right], \label{eq:H_sub} \\
H_\text{mix}&= \sum_{n=1,\sigma}^{N} V_{n}\left(d_{n,\sigma,0}^\dagger c_{n,\sigma} + \text{H.c.} \right), \label{eq:H_mix}
\end{align}
where $H_\text{SSH}$ is the SSH  Hamiltonian describing an isolated tPA chain with $N$ sites (i.e., -CH groups). 
The second-quantized fermionic operator $c_{n,\sigma}^\dagger$ ($c_{n,\sigma}$) creates (annihilates) an electron at site $n$ with spin projection $\sigma= \{\uparrow,\downarrow\}$. 
The first line of Eq. (\ref{eq:H_SSH}) describes the hopping processes of electrons along the chain, with $t_0$ the usual hopping parameter and $\alpha$ the electron-lattice coupling. The (classical) variable $u_n$ is the lattice degree of freedom corresponding to the displacement of the $n-$th group -CH along the axis of the molecule with respect to its equilibrium position. The second line in Eq. (\ref{eq:H_SSH}) is the elastic energy of the lattice, which penalizes distortions in the configuration of the nuclei. For simplicity, in this work we consider periodic boundary conditions, i.e., $c^\dagger_{n+N,\sigma}=c^\dagger_{n,\sigma}$. 

We now note that in most of the theoretical treatments of the SSH model the lattice degrees of freedom $u_n$ are usually assumed to be frozen in one of the Peierls-dimerized configurations, i.e., $u_n = \pm (-1)^n u_0$, where $u_0$ the amplitude of the Peierls distortion which minimizes the total ground-state energy. Within this assumption, the electron-lattice coupling in Eq. (\ref{eq:H_SSH}) takes the usual form $t_0 + \alpha \left(u_{n+1}-u_n\right)=t_0 + (-1)^n \delta t$, with $|\delta t |= |2\alpha u_0|$ the magnitude of the oscillating hopping term \cite{Asboth16_Short_course_on_TIs}. However, this hypothesis applies strictly to the isolated case, and its uncritical extension to more general situations, such as the one considered here, is not in principle justified. One of the goals in the present work is to obtain the equilibrium lattice configuration (i.e., the full vector $\{u_n\}$) by minimizing the total energy of the system \textit{in the presence} of the substrate at temperature $T=0$, a fact that drastically changes the physics of the Peierls dimerization, as we show in Sec. \ref{sec:results}.

The term $H_\text{sub}$ represents the presence of a metallic substrate, which is modeled as a collection of independent local reservoirs represented by semi-infinite tight-binding chains coupled to each site of the SSH chain. The degrees of freedom in the reservoirs are represented by the fermionic operators $d^\dagger_{n,\sigma,j}$ ($d_{n,\sigma,j}$) describing the creation (annihilation) of an electron with spin projection $\sigma$ at the $j$-th site of the $n$-th reservoir. While this  ``local bath'' model cannot account for spatial correlations in the substrate, it is in fact a reasonably good description of environment-induced phenomena, such as broadening of single-particle levels and quantum dissipation, of critical importance for our work. Here, the chemical potential $\mu_r$ and the reservoir hopping parameter $t_r$ can be chosen to describe different experimentally relevant situations.

Finally, the term $H_\text{mix}$ couples the site $n$ in the tPA molecule to the first site $j=0$ of the corresponding semi-infinite chain via the coupling parameter $V_n$. We assume this parameter to be site-dependent in order to account for possible inhomogeneities of the metallic surface. 

Our next step is the integration of the reservoir degrees of freedom. Technically, this can be achieved in the functional integral formalism \cite{negele}, which allows to express the partition function of the system as
\begin{align}
Z&=\int \mathcal{D}\left[\bar{c},c\right]\mathcal{D}\left[\bar{d},d\right] e^{-S_\text{SSH}-S_\text{sub}-S_\text{mix}},\label{eq:Z}
\end{align}
where the functional integral is expressed in terms of the Grassmann variables in imaginary time $\bar{c}_{n,\sigma}(\tau),c_{n,\sigma}(\tau)$ and $\bar{d}_{n,\sigma,j}(\tau), d_{n,\sigma,j}(\tau)$, corresponding to the degrees of freedom in both SSH chain and reservoirs, respectively. The corresponding Euclidean actions are defined as:
\begin{align}
S_\text{SSH}&=\int_0^\beta d\tau \sum_{n=1,\sigma}^{N} \bar{c}_{n,\sigma}(\tau)\partial_\tau c_{n,\sigma}(\tau)+ \int_0^\beta  H_\text{SSH}(\tau), \label{eq:S_SSH}\\
S_\text{sub}&=\int_0^\beta d\tau \sum_{n=1}^{N} \sum_{j=1,\sigma}^\infty \bar{d}_{n,\sigma,j}(\tau)\partial_\tau d_{n,\sigma,j}(\tau)+ \int_0^\beta  H_\text{sub}(\tau), \label{eq:S_sub}\\
S_\text{mix}&= \int_0^\beta  H_\text{mix}(\tau), \label{eq:S_mix}
\end{align}
where the Hamiltonians in these expressions have been expressed in the Heisenberg picture in imaginary time, and where we have defined $\beta=1/k_B T$ \cite{negele}.

The core of our approach consists in the derivation of an effective action for the SSH chain performing a partial integration over the substrate degrees of freedom. To that end, we introduce the partition function of the decoupled substrate, i.e., $Z^{(0)}_\text{sub}=\int \mathcal{D}\left[\bar{d},d\right] e^{-S_\text{sub}}$ and the ensemble average of an arbitrary operator $\mathcal{O}$ with respect to the reservoir degrees of freedom, i.e., $\langle \mathcal{O} \rangle_\text{sub}=\int \mathcal{D}\left[\bar{d},d\right] \mathcal{O} e^{-S_\text{sub}}/Z^{(0)}_\text{sub}$. With these definitions, we can express Eq. (\ref{eq:Z}) as
\begin{align}
Z&=Z^{(0)}_\text{sub}\int \mathcal{D}\left[\bar{c},c\right] e^{-S_\text{eff}},\label{eq:Z_integration_subs}
\end{align}
where we have defined the effective action of the SSH chain coupled to the substrate as
\begin{align}
S_\text{eff}&=S_\text{SSH}- \ln{\langle e^{-S_\text{mix}} \rangle_\text{sub}}. \label{eq:S_eff}
\end{align}

Using the Matsubara frequency representation \cite{negele}
\begin{align}
c_{n,\sigma}(\tau)&=\sum_{m} e^{-i\omega_m\tau} c_{n,\sigma}(i\omega_m),\\
d_{n,\sigma,j}(\tau)&=\sum_{m} e^{-i\omega_m\tau} d_{n,\sigma,j}(i\omega_m),
\end{align}
where $\omega_m=\frac{2\pi}{\beta}\left(m+\frac{1}{2}\right)$ are the fermionic Matsubara frequencies with integer number $m$, the effective action of the SSH chain after the integration of the reservoir degrees of freedom can be exactly expressed as
\begin{align}
    S_\text{eff}&=\beta \sum_m \left[\sum_{n=1,\sigma}^N \left(-i\omega_m -V_n^2 g_{00}(i \omega_m)\right) \bar{c}_{n,\sigma}(i\omega_m)c_{n,\sigma}(i\omega_m)\right.\nonumber \\
    &+\left. \sum_{n=1,\sigma}^{N-1} \left(-t_0-\alpha \left(u_{n+1}-u_n\right)\right)\left(\bar{c}_{n+1,\sigma}(i\omega_m) c_{n,\sigma}(i\omega_m)+ \text{H.c.}\right)\right] \nonumber \\
&+\frac{\beta K}{2}\sum_{n=1}^{N} \left(u_{n+1}-u_n\right)^2. \label{eq:S_eff2}
\end{align}
In the above expression, the reservoir degrees of freedom are encoded in the term 
$V_n^{2}g_{00}(i\omega_m)$, where the function $g_{jj^\prime}(i\omega_m)$ is the bare propagator of the electrons in the reservoir, i.e.,  
\begin{align}
    g_{jj^\prime}(i\omega_m)&=-\delta_{n,n^\prime} \delta_{\sigma,\sigma^\prime} \nonumber \\
    &\times \int_0^\beta d\tau \  e^{-i\omega_m \tau}\langle T_\tau d_{n,\sigma,j}(\tau)\bar{d}_{n^\prime,j^\prime,\sigma^\prime}(0)\rangle_\text{sub}, \label{eq:gjj_reservoir}
\end{align}
and where for compactness we have omitted spin and SSH-chain indices on the left hand-side since the reservoirs are all identical and SU(2)-symmetric. 

We stress that expression Eq. (\ref{eq:S_eff2}), where the substrate has been integrated out, is exact. 
Despite its simplicity, the term $V_n^{2}g_{00}(i\omega_m)$  introduces quantum dissipation effects, which manifests in renormalization and broadening of the electronic levels, crucially modifying the nature of the isolated one-dimensional SSH chain. An attractive feature of
this expression is that the case of the isolated SSH chain can be explicitly recovered 
by taking the limit $V_n \to 0$. 
Note that the simplicity of this expression is originated in the local nature of the metallic reservoirs (``local bath approximation''). A more realistic description of the substrate (e.g., a 2D or 3D metallic bath) is of course possible and would introduce indirect non-local correlations in the SSH chain, mediated by the metallic conduction electrons. However, a perturbative scaling dimension analysis shows that non-local, static correlation effects are subdominant, and the most relevant effects introduced by the substrate are local imaginary-time correlations \cite{lobos09_dissipation_scwires}.
We note that the local bath approximation is a useful approximation which has been implemented before in different contexts, e.g., in the case of few Kondo or Anderson impurities \cite{Garst04_QPT_of_Ising-coupled_Kondo_impurities, Zarand_Quantum_criticality_in_double_QDs, Mross09_2_Kondo_impurity_model_with_SOC}, Kondo lattice systems
\cite{Kojimani18_FM_QCP_in_KLModel,
Komijani19_Kondo_breakdown_1D_spin_chain}
, and dissipative quantum spin chains
\cite{Lobos12_Dissipative_XY_chain,  Lobos13_FMchains, Majumdar23_Dissipation_induced_localization_in_XXZ_chains}.

Using the Dyson's equation \cite{economou}, the expression of the propagator Eq. (\ref{eq:gjj_reservoir}) at $j=0$ can be easily computed  as 
\begin{align}
    g_{00}(i\omega_m)&= \frac{i\omega_m -\mu_r \pm \sqrt{(i\omega_m -\mu_r)^2-4t_r^2}}{2t^2_r}.\label{eq:g00_exact}
\end{align}
For simplicity, in what follows we assume the case of half-filling both in the substrate and the SSH chain, which implies that the chemical potential of the reservoirs must be set to $\mu_r=0$. Of course, this is not a generic situation in our model. However, since our interest is the description of the Peierls-dimerized phase of the isolated SSH chain (and its possible destabilization due to the coupling to the substrate), the condition $\mu_r=0$, which corresponds to half-filling in the SSH chain, ensures the perfect nesting of the one-dimensional Fermi surface which triggers the Peierls instability due to the existence of a lattice phonon with momentum $q=2 k_F=\pi/a$ connecting the two Fermi points $k_F=\pm  \pi/2a$ \cite{Heeger88_Solitons_in_conducting_polymers}. 
In contrast, whenever 
the chemical potential in the substrate is $\mu_r \neq 0$, charge-transfer effects between the substrate and the SSH chain modify the band filling of the SSH chain and frustrate the perfect nesting of Fermi surface and the Peierls dimerization phenomenon. Thus, a convenient way to exclude other interfering effects which would complicate our analysis is to set $\mu_r=0$. However, we stress that our model is more general and could potentially describe other experimentally relevant situations. 

As a last point in this Section, we note that under typical experimental conditions the bandwidth of the metal is much larger than the bandwidth of the SSH chain (i.e., $t_r \gg t_0$), a fact that enables to  take the wide bandwidth limit (WBL), and to replace the expression Eq. (\ref{eq:g00_exact}) by $g_{00}(i\omega_m)\simeq -i\text{sgn}(\omega_m) \pi \rho_0$, where $\rho_0$ is the density of states of the reservoir at the Fermi energy $\rho_0 =-g^\text{ret}_{00}(0+ i 0^+)/\pi$ obtained by analytic continuation from Eq. (\ref{eq:g00_exact}). As discussed in the next Section, this simplifying approximation allows for a more direct numerical determination of the energy minimum and yields analytical expressions for the ground-state energy.

\section{Numerical minimization of the total energy}\label{sec:minimization}

So far, we have not discussed how the lattice degrees of freedom, treated here as free variational parameters in our theoretical approach, are determined. As mentioned in Section \ref{sec:intro}, in order to obtain physically consistent results, the equilibrium positions of the -CH groups should emerge from the global minimization of the total energy of the system. From Eqs. (\ref{eq:Z_integration_subs}) and (\ref{eq:S_eff}) we can write the Helmholtz free energy of the total system as $F=-\beta^{-1}\ln{Z}=F^{(0)}_\text{sub}+F_\text{eff}$, where $F^{(0)}_\text{sub}$ is a (formally infinite) constant contribution from the decoupled substrate, and $F_\text{eff}$ is the contribution of the coupled SSH chain. To numerically compute this last contribution to the energy, we assume a given (a priori random) distribution of the positions $\{u_n\}$. The effective action Eq. (\ref{eq:S_eff2}) can be compactly expressed in terms of the vectors $\Psi_\sigma (i \omega_m)=\{c_{1,\sigma}(i\omega_m), c_{2,\sigma}(i\omega_m), \dots, c_{N,\sigma}(i\omega_m)\}^T$ and $\mathbf{u}=\{u_1, u_2, \dots, u_N\}$ as
\begin{align}
S_\text{eff}&=\beta \sum_m \sum_\sigma \bar{\Psi}_{\sigma}(i\omega_m) \mathcal{G}^{-1}(i\omega_m) \Psi_{\sigma}(i\omega_m) \nonumber \\
&+\frac{\beta K}{2} \sum_{n=1}^{N} (u_{n+1}-u_n)^2, \label{eq:S_eff_matrix}
\end{align}
where we have defined the $N \times N$ propagator matrix as
\begin{align}\label{eq:inverse_G_matrix}
\mathcal{G}^{-1}(i\omega_m)&=\mathcal{H}^\text{(0)}_\text{SSH}(\{\mathbf{u}\}) -i  \Gamma(\omega_m) 
\end{align}
where $\mathcal{H}^\text{(0)}_\text{SSH}(\{\mathbf{u}\})$ is the matrix corresponding to the electronic part of the isolated SSH chain Eq. (\ref{eq:H_SSH}), and  $\Gamma(\omega_m)$ is a diagonal matrix representing the effect of the substrate,  defined as $\left[\Gamma(\omega_m)\right]_{n,n^\prime}=\delta_{n,n^\prime} \text{sign}(\omega_m) \gamma_n$, where we have defined the local broadening parameter $\gamma_n\equiv \pi V_n^2 \rho_0 $. 

The free energy can be formally obtained from Eq. (\ref{eq:S_eff_matrix}) as \cite{negele}
\begin{align}
    F_\text{eff}&=-\frac{1}{\beta}\ln{\text{det}}\left[\mathcal{G}^{-1}\right] + \frac{K}{2} \sum_{n=1}^{N} (u_{n+1}-u_n)^2.
\end{align}
In practice, to evaluate this expression, we first diagonalize the matrix $\mathcal{G}^{-1}(i\omega_m)$ using the bi-orthonormal eigenvector basis, i.e., 
\begin{align}
\left[
\mathcal{H}^\text{(0)}_\text{SSH}(\{\mathbf{u}\}) -i  \Gamma(\omega_m)
\right]
\chi_{\nu}(\omega_m)&=\lambda_{\nu}(\omega_m)\chi_{\nu}(\omega_m)\nonumber \\
\left[
\mathcal{H}^\text{(0)}_\text{SSH}(\{\mathbf{u}\})+i  \Gamma(\omega_m)
\right]
\chi^*_{\nu}(\omega_m)&=\lambda^*_{\nu}(\omega_m)\chi^*_{\nu}(\omega_m),\label{eq:eigenvalue_eqs}
\end{align}
where $\lambda_\nu (\omega_m) = \epsilon_\nu -i\text{sign}(\omega_m)\gamma_\nu$ is a complex eigenvalue whose real part corresponds to the frequency-independent  eigenenergy of the state $\chi_\nu$, whereas $\gamma_\nu = \left| \text{Im}\ \lambda_\nu  (\omega_m) \right|$ is its broadening \cite{Note_Dissipation_Article_Arancibia20206}. In the above expressions, all the effects of the coupling to the metallic substrate are encoded in both the eigenvectors $\chi_\nu$ and eigenvalues $\lambda_\nu$. Note that the WBL is crucial here to simplify the frequency $\omega_m$-dependence of the dissipative term, and to enable an appropriate eigenvalue equation  describing the effects introduced by the metallic environment.
In terms of these eigenvalues, the free energy can be expressed as a conventional Matsubara sum 
\begin{align}
    F_\text{eff}&=-\frac{1}{\beta}\sum_{m=-\infty}^{\infty} \sum_{\nu=1}^N \ln{\left[\beta\left(i\omega_m - \lambda_\nu (\omega_m) \right) \right]} \nonumber \\
    &+ \frac{K}{2} \sum_{n=1}^{N} (u_{n+1}-u_n)^2.
\end{align}
Performing the Matsubara sum as an integral along the cut (i.e., real frequency axis) yields the expression in the limit $T\rightarrow 0$
\begin{align}
    E_\text{GS}&=\frac{1}{\pi}\sum_{\nu=1}^N \left[-\frac{\pi (W+\mu_r)}{2}+\frac{\gamma_\nu}{2}\ln{\left(
    \frac{\gamma_\nu^2+(\epsilon_\nu-\mu_r)^2}{\gamma_\nu^2+(\epsilon_\nu+W)^2}\right)}\right.\nonumber \\
    &+\left. \left(\epsilon_\nu+W \right)\arctan{\left(\frac{\epsilon_\nu+W}{\gamma_\nu} \right)} - \left(\epsilon_\nu-\mu_r\right)\arctan{\left(\frac{\epsilon_\nu-\mu_r}{\gamma_\nu}\right)}\right]\nonumber \\
    &+ \frac{K}{2} \sum_{n=1}^{N} (u_{n+1}-u_n)^2.
\end{align}
Since this expression has been evaluated assuming a given configuration of the lattice $\mathbf{u}$, the total ground-state energy is $E_\text{GS}\equiv E_\text{GS}(\mathbf{u})$. 
The minimum of the energy is then searched in the $N$-dimensional space spanned by the individual parameters $u_n$. To that end, we use a numerical routine implementing the Broyden-Fletcher-Goldfarb-Shanno (BFGS) algorithm \cite{pang2006introduction}, using a tolerance parameter $\sqrt{(\mathbf{u}^{(k)}-\mathbf{u}^{(k-1)})^2}\leq 10^{-4} a$ between iterations $k$ and $k-1$. To accelerate convergence,  we use an initially dimerized seed $\mathbf{u}^{0}$. 

As we show explicitly in the next Section, identifying the global minimum of the energy in the 
$N-$dimensional space spanned by the lattice configuration vector 
$\mathbf{u}$ is a crucial step in properly accounting for environmental effects. We stress that this step is typically neglected in most of the theoretical treatments of the SSH model, an approximation that is only justified when the tPA chain is considered in isolation. However, in the presence of an external coupling to the environment or when the tPA chain is contacted in electrical circuits (see, e.g., Ref. \cite{Arancibia25_Towards_electrical_DW_control}), this assumption must be revisited.

\section{Results}\label{sec:results}
\subsection{SSH chain coupled to a homogeneous metallic surface}\label{sec:homogeneous}

\begin{figure}[t]
 \includegraphics[width=0.95\columnwidth]{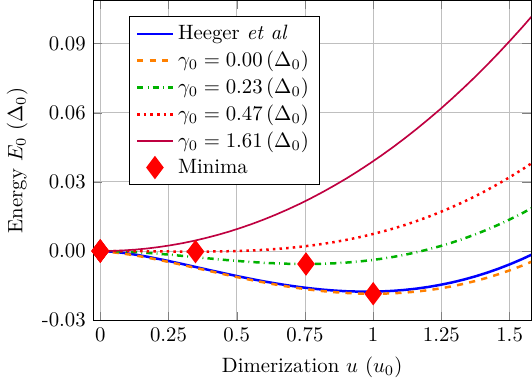}
\caption{\label{fig:EGS_minima}(Color online) Total energy of the system at $T=0$ as a function of the dimerization parameter $u$ for different values of the broadening $\gamma_0$. Beyond a critical value of $\gamma_0$ the Peierls-dimerized state (with $u_0 \neq 0$) is destabilized and the undistorted, metallic state with $u_0=0$ is favoured.}
\label{system}
\end{figure}

\begin{figure}[t]
\includegraphics[width=0.9\columnwidth]{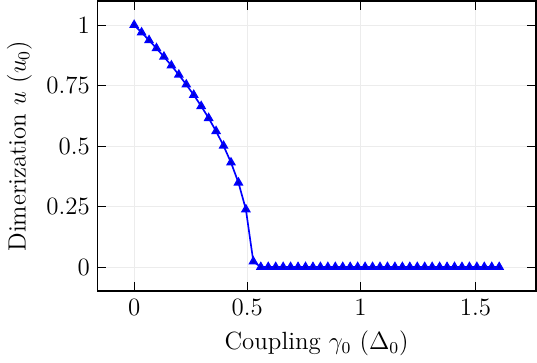}
\caption{\label{fig:u0_vs_gamma}(Color online) Dimerization parameter $u_0$ vs the coupling parameter $\gamma_0$ for a tPA chain deposited ontop of a uniform metallic substrate. The value of the Peierls distortion $u_0$ is monotonically reduced as $\gamma_0$ increases, and beyond a critical value $\gamma_{0, \text{cr}}\approx 0.46 \Delta_0$, the system becomes metallic.}
\label{system}
\end{figure}

We first implement our approach to the case of a tPA molecule deposited ontop of a uniform and homogenous metallic substrate. Within our theoretical framework, this case is described by a uniform coupling parameter $V_n=V_0$ and, therefore, a uniform broadening $\gamma_0=\pi V_0^2 \rho_0$. In this case,  the dissipation matrix reduces to  $\Gamma(\omega_m)=\gamma_0 \text{sign}(\omega_m) \mathbb{1}_{N \times N}$, and the eigenvectors of the non-hermitian operator $\mathcal{H}^\text{(0)}_\text{SSH}(\{\mathbf{u}\})+i  \Gamma(\omega_m)$ are identical to those of $\mathcal{H}^\text{(0)}_\text{SSH}(\{\mathbf{u}\})$ (i.e., the uncoupled SSH chain). The eigenvalues are trivially expressed as $\lambda_\nu (\omega_m) = \epsilon^{(0)}_\nu - i\text{sign}(\omega_m) \gamma_0$, with $\epsilon^{(0)}_\nu$ the eigenvalues corresponding to the isolated SSH chain. Imposing periodic boundary conditions, the ground-state solution corresponds to a Peierls-dimerized system with $u_n=(-1)^n u$, with a generic distortion parameter $u$. Under these conditions, the  minimization problem greatly simplifies and effectively consists in finding the optimal distortion $u\rightarrow u_0$ which minimizes the total energy $E_\text{GS}(u)$. In Fig. \ref{fig:EGS_minima} we show the evolution of the total ground-state energy as a function of $u$ for different values of $\gamma_0$. For clarity, we have only plotted the difference $\Delta E_\text{GS}(u,\gamma_0)=E_\text{GS}(u,\gamma_0)-E_\text{GS}(0,\gamma_0)$ where the value at $u=0$ have been substracted in all curves. For each value of $\gamma_0$, the red diamonds show the position of the minima found with the numerical procedure described in Section \ref{sec:minimization}. As is shown in Fig. (\ref{fig:EGS_minima}), the Peierls-distorted ground state is continuously destabilized as $\gamma_0$ increases, and beyond a critical coupling $\gamma_{0,\text{cr}}\approx 0.47 \Delta_0$, 
where $\Delta_0$ is the Peierls gap for the uncoupled SSH chain, the dimerization is completely suppressed and the metallic state for the SSH chain becomes the new ground state. Note that this destabilization of the Peierls-dimerized state is a pure quantum dissipation phenomenon introduced by the metallic substrate, which emerges as a direct consequence of the self-consistent treatment of both electronic and lattice degrees of freedom. This ``metallization'' transition in the SSH chain can be seen more clearly in Fig. \ref{fig:u0_vs_gamma}, where we show the different minima $u_0$ occurring in the system as function of $\gamma_0$ (i.e., red diamonds in Fig. (\ref{fig:EGS_minima})). This plot qualitatively resembles the mean-field ferromagnetic-paramagnetic 
second-order phase transition of the Ising model, where $u_0$ plays the role 
of the magnetization and $\gamma_0$ plays the role of the temperature. Indeed, our results correspond to the mean-field limit of a more general theoretical framework in which the lattice displacements  are true quantum-mechanical operators $\hat{u}_n$ canonically conjugate to momenta $\hat{p}_n$. In this picture, our classical variational parameters $u_n$ correspond to their expectation values, i.e., $u_n = \langle \hat{u}_n \rangle$, and therefore the homogeneous parameter $u$ in Fig. \ref{fig:EGS_minima} can  be regarded as the global order parameter of the transition.
Physically, this can be interpreted as the result of a frustrated Fermi-surface nesting due to single-particle lifetime effects. From a different perspective, the dissipation effects introduced by the substrate are a natural consequence of coupling the SSH chain to a higher-dimensional system, where the conditions for the Peierls theorem are not fulfilled (i.e., the complete system is not one-dimensional) \cite{Peierls_Quantum_Theory_of_Solids}.  

We believe these results are relevant for the correct interpretation of ongoing experiments, as those of Ref. \cite{Wang19_Solitons_in_individual_PA_molecules}, where tPA molecules were grown on top of Cu(110) surface. Using angle-resolved photoemission spectroscopy
(ARPES), the authors identified metallic behavior in regions
where the tPA molecules couple strongly to the metallic
substrate, in contrast to the semiconducting character
observed on the oxidized CuO terraces. We discuss this in more detail in Sec. \ref{sec:summary}.

\subsection{SSH chain coupled to a non-uniform metallic-insulating substrate}

Of particular interest to ongoing experiments is the case of a SSH chain coupled to a inhomogeneous surface having both metallic and insulating regions, as depicted in Fig. \ref{fig:onsurface_tPA_scheme}(a). In the abovementioned  Ref. \cite{Wang19_Solitons_in_individual_PA_molecules},  this is the case of tPA molecules bridging the Cu/CuO interface. In that case, spatially-resolved STM results show 
that the two phases described in the previous Section, dimerized and undimerized, can occur simultaneously within a single tPA molecule:
while the Peierls-dimerized phase can locally occur in the region ontop the CuO terrace, the undistorted phase can occur in the region grown on top the bare Cu. These experimental findings remain to be understood and call for a proper theoretical interpretation, particularly regarding their implications for the existence (or absence) of topological defects and solitons in real physical systems.

To address this experimental situation, we focus on a $N=200$ site chain with periodic boundary conditions, and we model the Cu/CuO terrace with a spatially varying coupling $V_n$ such that $V_{n} = V_0$ for $n\in[50,150]$, where $V_0$ is an adjustable parameter giving rise to a local broadening parameter $\gamma=\pi V_0 \rho_0$ on those sites, and $V_{n}=0$ elsewhere (representing the region of the chain above the decoupling layer of CuO). We then proceed as described in Sec. \ref{sec:minimization} and search for the configuration of lattice positions $\{u_n\}$ that minimizes the total energy in the presence of the inhomogeneous coupling $V_n$. However, instead of searching directly in $u_n$-space, we introduce the vector of \textit{differences} of neighboring positions
\begin{align}
y_{n}&= u_{n+1}-u_{n}.\label{eq:yn}
\end{align}
Although mathematically equivalent, this quantity is slightly more convenient from the numerical perspective  as it enters directly in the Hamiltonian Eq. (\ref{eq:H_SSH}). 

Due to the rapid oscillatory behavior of the Peierls-dimerized component, which complicates the visualization of the data, a direct plot of the configuration $\{y_n\}$ obtained at the energy minimum is not very illuminating. To express our results more transparently, we introduce the Fourier decomposition of $y_n$
\begin{align}
y_n&=\sum_{q=-\pi/a}^{\pi/a} \frac{e^{i q a n}}{\sqrt{N}} y_q, 
\end{align}
where the Fourier components are $q=2\pi l/Na$, with $l$  taking integer  values in the interval $l \in [-\frac{N}{2}, \dots, \frac{N}{2})$.  
Next, once all the Fourier components $y_q$ have been obtained (e.g., by the means of the FFT algorithm), we can simplify the visualization of our results by separating the ``slow'' from the ``fast'' modes. This can be achieved by defining the quantities
\begin{align}
    S_n &= \sum_{|q|<\pi/2a} \frac{e^{i q a n}}{\sqrt{N}} y_q, \\
    R_n &=\sum_{|q|>\pi/2a} \frac{e^{i q a n}}{\sqrt{N}} y_q.
\end{align}
Note that these two components of the lattice distortion are associated to physically different mechanisms: since  the slow component $S_n$ consists in Fourier modes which are strongly peaked around $q\sim 0$, it describes 
smooth configurations of the lattice difference-field $y_n$, such as smooth contractions or expansions of the chain. On the other hand, the rapid component $R_n$ consists of Fourier modes which are peaked around $q\sim \pi/a$, and therefore are physically related to the mechanism of Peierls dimerization. Once the rapidly oscillating component has been isolated, for visualization purposes we can define the staggered component $\bar{R}_n=(-1)^n R_n$, which allows a better insight into the Peierls-dimerized component. We stress that no information is lost in the above decomposition, since the original configuration vector $y_n$ can be fully recovered by simply adding $S_n+(-1)^n \bar{R}_n$. As a practical example, let us now think of the case of a completely isolated tPA chain. As we have already mentioned, in this case the configuration that minimizes the energy is the dimerized pattern $u_n = -(-1)^n u_0$, which results in $y_n =(-1)^n 2u_0$. This means that, for this particular case, the state of the lattice is fully described by a \textit{single} Fourier component at $q= \pi/a$, corresponding to $y_{\pi/a}=2 u_0$.

\begin{figure}
    \centering
    \includegraphics[width=1\linewidth]{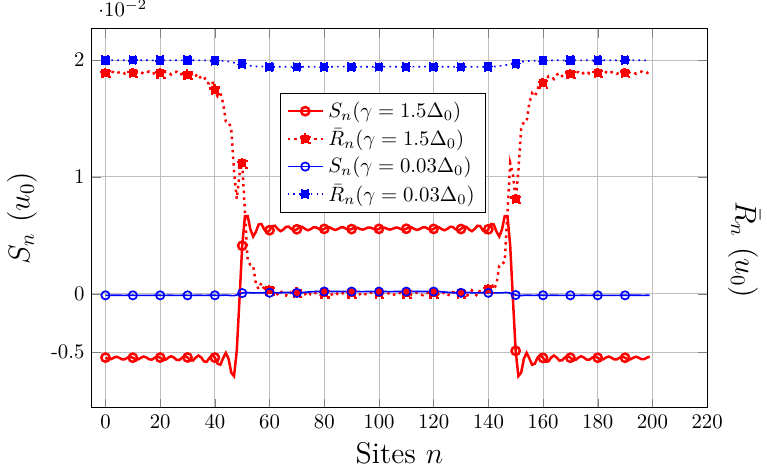}
    \caption{(Color online) Profile of the smooth ($S_n$) and staggered ($\bar{R}_n$) components of the lattice distortion $y_n$ [see Eq. (\ref{eq:yn})] for two different values of the coupling $\gamma=0.03 \Delta_0$ and $\gamma=1.5 \Delta_0$ (blue and red symbols, respectively), applied at sites $n \in [50,150]$ (i.e., the central segment). In the weak-coupling case $\gamma=0.03 \Delta_0$, both $S_n$ and $\bar{R}_n$ remain essentially unchanged compared to the isolated SSH model. In contrast, in the strong-coupling case $\gamma=1.5 \Delta_0$, both quantities differ significantly from their isolated-case values. In particular, the Peierls dimerization characteristic of the isolated SSH model is destabilized in the coupled region, and the molecule becomes locally metallic.
}
    \label{fig:lattice_configurations_inhomogeneous_gamma}
\end{figure}

Let us turn now to the more involved case of a non-homogeneous coupling $V_n$, where both lattice-distortion components $S_n$ and $\bar{R}_n$ appear. In Fig.~\ref{fig:lattice_configurations_inhomogeneous_gamma} we show these two components for two different values of the local broadening in the metallic region, namely $\gamma=0.03 \Delta_0$ and $\gamma=1.5 \Delta_0$. These values are chosen such that one lies below the critical homogeneous value $\gamma_{0,\text{cr}}=0.46 \Delta_0$ found in the previous Section, while the other lies above it. Let us first analyze the case $\gamma=0.03 \Delta_0$, which represents weak coupling to the metallic region (middle region in Fig.~\ref{fig:lattice_configurations_inhomogeneous_gamma}). Note that both $S_n$ and $\bar{R}_n$ (shown as blue circles and blue squares) remain essentially unperturbed by the presence of the metal, which induces only small renormalizations with respect to the insulating region. In particular, the fast component $\bar{R}_n$ fluctuates around the value $2u_0$, corresponding to the completely isolated case discussed previously. By contrast, in the strongly coupled case $\gamma=1.5 \Delta_0$, the situation is strikingly different. The $\bar{R}_n$ component (shown as red squares) saturates at $\sim 2u_0$ for sites above the insulating region, indicating that its Peierls-dimerized structure survives locally, but drops to zero in the segment directly coupled to the metal, signaling the complete destruction of the dimerized pattern in that region. We speculate that the tPA molecules in Ref.~\cite{Wang19_Solitons_in_individual_PA_molecules} operate precisely in this regime. In addition, the $S_n$ component, shown as red circles in Fig.~\ref{fig:lattice_configurations_inhomogeneous_gamma}, reveals an expansion of the ionic structure on top of the metallic region, highlighting the importance of allowing the lattice to relax in order to minimize the energy. Since in our calculations the total length of the system is kept constant (i.e., the constraint $\sum_n S_n = 0$ is imposed), the lattice structure exhibits a contraction above the insulating region that compensates for the expansion induced by the metallic environment. We stress, however, that this constraint would not be present for a free-ended molecule, and we speculate that in that case only the metal-induced expansion would prevail.

To connect the structural modifications discussed above with their electronic signatures, we now analyze the local density of states (LDOS), 
\begin{align}
\rho_{n}(\omega) 
&= -\frac{1}{\pi} \text{Im}\left[\mathcal{G}^\text{ret}(\omega)\right]_{n,n} 
\label{eq:LDOS}
\end{align}
where the matrix $\mathcal{G}^\text{ret}(\omega)$ is obtained from Eq. (\ref{eq:inverse_G_matrix}) by analytic continuation to real frequencies $i \omega_m \to \omega + i 0^+$. 
The numerical evaluation of $\mathcal{G}^\text{ret}(\omega)$ is technically done exploiting the eigenmode expansion given by the eigenvalue Eqs. (\ref{eq:eigenvalue_eqs}), which for complex frequencies above the cut (real frequency axis) simplify to   
\begin{align}
\left[
\mathcal{H}^\text{(0)}_\text{SSH}(\{\mathbf{u}_\text{eq}\}) -i  \Gamma
\right]
\chi_{\nu}&=\lambda_{\nu}\chi_{\nu}\nonumber \\
\left[
\mathcal{H}^\text{(0)}_\text{SSH}(\{\mathbf{u}_\text{eq}\})+i  \Gamma
\right]
\chi^*_{\nu}&=\lambda^*_{\nu}\chi^*_{\nu},\label{eq:eigenvalue_eqs_retarded_G}
\end{align}
where the equilibrium configuration $\mathbf{u}_\text{eq}$ is used, and where matrix $\Gamma$ is now frequency-independent. Then, the retarded Green's function can be expressed as:
\begin{align}
\left[\mathcal{G}^\text{ret}(\omega)\right]_{n,n^\prime}&=\sum_\nu
\frac{\chi_\nu(n)\chi_\nu(n^\prime)}{\omega  -\epsilon_\nu +i\gamma_\nu},
\end{align}

\begin{figure}
    \centering
    \includegraphics[width=1\linewidth]{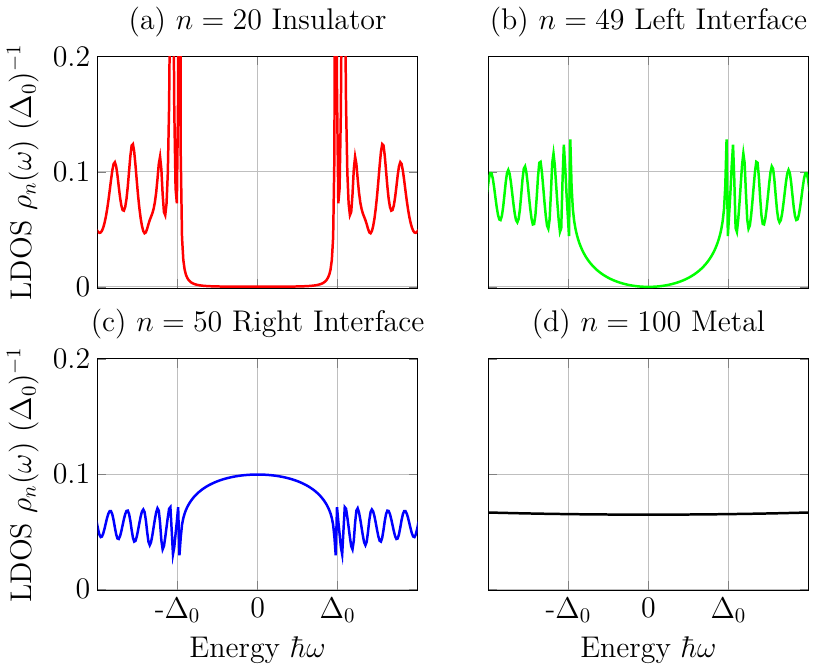}
    \caption{(Color online) Local density of states (LDOS) $\rho_{n}(\omega)$ computed at four representative positions along a chain with inhomogeneous coupling $V_n$. Panel (a) shows the LDOS at $n=20$ (insulating region), panel (b) at $n=49$ (interface on the insulating side), panel (c) at $n=50$ (interface on the metallic side), and panel (d) at $n=100$ (metallic region).}
    \label{fig:LDOS}
\end{figure}

 The LDOS provides a direct probe of the electronic structure, as it is proportional to the differential conductance measured in STM experiments, $\rho_{n}(\omega)\propto dI/dV$ ~\cite{Tinkham_Introduction_to_superconductivity}. Then, the LDOS obtained from Eq.~\eqref{eq:LDOS} can be directly compared with the experimental observations reported in Ref.~\cite{Wang19_Solitons_in_individual_PA_molecules}. In  Fig.~\ref{fig:LDOS} we display the LDOS $\rho_{n}(\omega)$, computed from Eq. (\ref{eq:LDOS}) for a coupling $\gamma_{0}=1.6\,\Delta_{0}$, at representative positions along the chain: a) site $n=20$ (deep inside the insulating  region, b)
site $n=49$ (interface, insulating side), c) site $n=50$ (interface, metallic side ), and site $n=100$ (center of the metallic region). In panel a) the LDOS exhibits the characteristic features of a semiconductor, including an energy gap and Van~Hove singularities at the band edges. On the other hand, in panel d) the LDOS displays a featureless flat profile typical of a metallic spectrum. These results are the expected deep inside the insulating or metallic side, respectively, and reproduce the experimental STM differential-conductance results in Ref. \cite{Wang19_Solitons_in_individual_PA_molecules}. 

\begin{figure}
    \centering
    \includegraphics[width=1\linewidth]{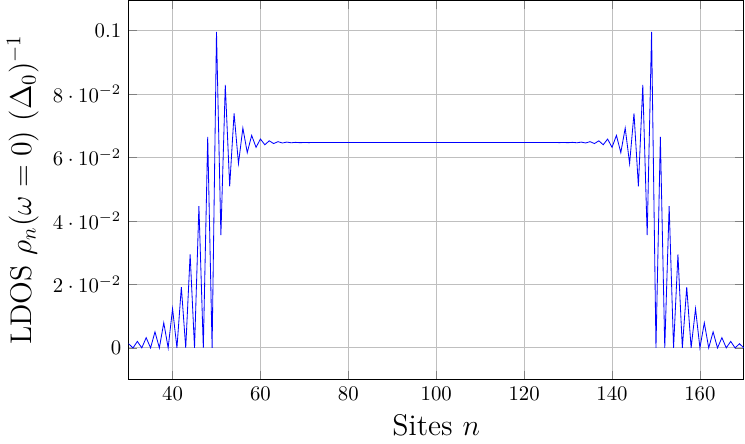}
    \caption{(Color online) Spatial profile of the LDOS $\rho_{n}(\omega=0)$ showing its progressive decay away from the interface. The strong oscillations reveal interference patterns compatible with Friedel oscillations (see discussion in the maintext).
}
    \label{fig:DOS}
\end{figure}

On the other hand, the interfacial behavior displays features that diverge from the experimental interpretation of Ref.~\cite{Wang19_Solitons_in_individual_PA_molecules}, where a soliton-like midgap state was inferred. Our theoretical analysis instead reveals a pattern consisting of LDOS in-gap  depletions on even sites and  LDOS enhancements on odd sites, with amplitudes that decay smoothly away from the interface. This behavior, summarized by the spatial profile of $\rho_{n}(\omega=0)$ in Fig. \ref{fig:DOS}, reflects an alternation and attenuation of oscillations consistent with the scattering pattern of extended electronic wave functions (i.e., Friedel oscillations modulated by a tunnel-effect evanescent profile), rather than the presence of a localized midgap excitation. From a different perspective, this conclusion is supported using a one-dimensional Dirac model with a position-dependent mass term to model the Peierls insulator–metal interface (i.e., the Jackiw-Rebbi model, see \cite{Jackiw76_Jackiw_Rebbi_soliton, Shen2012_Topological_Insulators}). In fact, assuming a mass $m_1\neq 0$ for $x<0$ (Peierls-dimerized region) and  $m_2=0$ for $x>0$ (metallic part), one can conclude that an electronic bound-state at $E=0$ cannot exist at the interface.

From an experimental standpoint, these findings suggest a refined interpretation of STM measurements in experimental realizations of the SSH model on inhomogeneous metallic surfaces. In particular, they indicate that resonant LDOS features may arise not only from individual topological states but also from collective interference effects generated by structural gradients and inhomogeneous coupling. This insight is pertinent to the design of quantum nanodevices—organic, hybrid, or atomic—in which local conductance can be tuned by engineering the contact morphology or the spatial profile of the coupling to the substrate. Within this framework, our formalism not only captures the electronic properties of tPA on composite substrates  but also provides a general tool for modeling and predicting electronic behavior in one–dimensional architectures featuring functional interfaces.

From an experimental standpoint, these results indicate that an inhomogeneous coupling to the substrate can induce structural gradients along the molecule, thereby modulating its local metallic or semiconducting character.

\section{Summary and conclusions}\label{sec:summary}
In this work we introduced an extended Su-Schrieffer-Heeger (SSH) model to examine the zero-temperature electronic and structural properties that emerge when a tPA chain couples to a metallic substrate. This is of particular interest to  recent experimental works where on-surface sythesis of organic polymer materials have been fabricated. We focused on two experimentally relevant configurations: the case of a homogeneous metallic substrate, in which all sites in the SSH chain couple to the substrate identically, and an inhomogeneous substrate, where a part of the chain is connected to a metallic region of the substrate whereas the rest is deposited ontop of an insulating decoupling layer.

In the homogeneous case, we found that increasing the chain–substrate coupling introduces broadening in the electronic levels of the SSH. Importantly, this broadening progressively suppresses the Peierls distortion due to the suppression of the Fermi surface nesting, and beyond a critical value  $\gamma_{0, \text{cr}}$, a dissipation-driven quantum phase transition occurs from a Peierls-dimerized insulator to an undistorted metallic state. This behavior is consistent with recent experimental ARPES and STM observations in tPA chains grown on a metallic surface \cite{Wang19_Solitons_in_individual_PA_molecules}. In terms of our model, these experimental results are well described by a SSH chain in the metalized phase, suggesting that the effective broadening introduced by the  metallic Cu(110) substrate is $\gamma > \gamma_{0,\text{cr}}$.

A theoretical study similar to ours is carried out in Ref.~\cite{edalatmanesh2025non}. In this work, the authors analyze a finite SSH chain with open boundary conditions coupled to a metallic reservoir, with the aim of investigating the possible emergence of in-gap states. Their results show that, under certain boundary conditions, such states indeed appear. However, in their treatment, as is common in other theoretical approaches discussed previously,  the chain is assumed to be frozen in a dimerized configuration and the coupling parameter $\gamma$ is taken to be independent of the ionic positions. The essential difference with our approach lies in the enforcement of self-consistency, which establishes a nontrivial relationship between the parameter $\gamma$ and the ionic configurations of the system [Eq.~(\ref{eq:eigenvalue_eqs})) together with the minimization procedure]. In this way, our model explicitly incorporates structural degrees of freedom, thereby enriching the physical description and enabling a more realistic comparison with experimental results.

For the inhomogeneous case, our results point to the possibility for the two phases found in homogeneous case to coexist in the same tPA molecule, producing marked structural and electronic differentiation along the chain. Local coupling to a metallic portion of the substrate induces a metallic phase in the SSH portion directly above, with an equidistant lattice configuration and the absence of a gap in the local density of states (LDOS). On the other hand, the portion of the SSH chain coupled to an insulating substrate locally preserves its dimerization and insulating properties, showing a well-defined gap in the LDOS. In addition, a small smooth expansion appears in the metallic SSH portion. 

These results directly apply to the correct interpretation of experimental results. In particular, in contrast to the interpretation of Ref. \cite{Wang19_Solitons_in_individual_PA_molecules}, which attributes the LDOS protrusions at the metal/insulator interface to a localized soliton-like state, our microscopic analysis indicates that these features arise from collective interference effects of  scattering wave functions, rather than from midgap localized excitations. In particular, we demonstrate that the spatial pattern shown in Fig.~\ref{fig:DOS}, qualitatively similar to the STM results found in Ref. \cite{Wang19_Solitons_in_individual_PA_molecules}, does not require invoking solitonic midgap states, but follows naturally from electronic oscillations induced by the spatial variation of the substrate coupling. These results challenge the idea that perfect lattice-dimerization is a universal feature in SSH systems, and highlight the need to consider fully relaxed configurations in which the ionic displacements $u_{n}$ adjust self–consistently.

From the viewpoint of nanoscale device design, the configurations analyzed in the work could be exploited to create active regions with electronically tunable properties through controlled gap engineering. Moreover, the theoretical framework developed here is readily extendable to other systems that can be described within the extended SSH model, including organic heterochains, one-dimensional atomic lattices, and hybrid devices in which environment–induced coupling plays a functional role in electronic transport.
Moreover, our results demonstrate that controlling the chain–substrate coupling not only enables tuning the transition between semiconducting and metallic regimes but also offers a handle to manipulate interference patterns effects in hybrid nanostructures. This understanding is crucial for guiding the development of devices based on conducting polymers and topological materials, where electronic properties can be engineered through atomic–scale control of the environment.

Finally, we mention that beyond the specific case of tPA, our findings provide a conceptual framework applicable to a broad class of quantum materials. The possibility of mapping the extended SSH model onto non–organic systems, such as those investigated in \cite{cheon2015chiral}, strengthens the relevance of our approach for the design of nanodevices in which topology, electronic structure, and the physical environment are intricately coupled.

\bibliographystyle{apsrev}

\begin{thebibliography}{30}
\bibitem[{Not()}]{Note_Dissipation_Article_Arancibia20206}
\bibinfo{note}{Note that, unlike the Hermitian case, the existence of a biorthonormal basis that diagonalizes a non-Hermitian matrix is not guaranteed in general. However, situations where such a basis fails to exist occur only at isolated points in parameter space (the so-called ``exceptional points'') and in practice the non-Hermitian operator in Eq. (\ref{eq:eigenvalue_eqs}) can be diagonalized in most of the cases. In all the results shown in this work we have numerically verified that the biorthonormal basis allowing diagonalization in fact exists.}
\end{thebibliography}

\end{document}